\documentclass[aps,prl,twocolumn]{revtex4-1}

\usepackage{amsmath,amssymb,amsfonts,graphicx,tabularx,xcolor}

\usepackage[unicode=true,colorlinks=true]{hyperref}

\hypersetup{linkcolor=blue,citecolor=blue,urlcolor=blue}

\newcommand{\comments}[1]{}
\newcommand{\mb}[1]{\mathbf{#1}}

\def\U{\mathrm{U}(1)}

\def\Z{\mathbb{Z}}

\begin{document}

\title{Continuous phase transitions between fractional quantum Hall states \\ and symmetry-protected topological states}

\author{Ying-Hai Wu}
\affiliation{School of Physics and Wuhan National High Magnetic Field Center, Huazhong University of Science and Technology, Wuhan 430074, China}

\author{Hong-Hao Tu}
\affiliation{Institut f\"ur Theoretische Physik, Technische Universit\"at Dresden, 01062 Dresden, Germany}

\author{Meng Cheng}
\affiliation{Department of Physics, Yale University, New Haven, Connecticut 06511-8499, USA}

\begin{abstract}
We study quantum phase transitions in Bose-Fermi mixtures driven by interspecies interaction in the quantum Hall regime. In the absence of such an interaction, the bosons and fermions form their respective fractional quantum Hall (FQH) states at certain filling factors. A symmetry-protected topological (SPT) state is identified as the ground state for strong interspecies interaction. The phase transitions between them are proposed to be described by Chern-Simons-Higgs field theories. For a simple microscopic Hamiltonian, we present numerical evidence for the existence of the SPT state and a continuous transition to the FQH state. It is also found that the entanglement entropy between the bosons and fermions exhibits scaling behavior in the vicinity of this transition.
\end{abstract}

\maketitle

{\em Introduction} --- The collective behavior of a large number of microscopic objects is a fascinating topic. In quantum condensed matter physics, one central task is to elucidate the possible phases and transitions between them for a given many-body system. A large class of phases and transitions is characterized by spontaneous breaking of global symmetries, described by the Landau-Ginzburg theory. However, quantum phases of matter beyond the symmetry-breaking framework have also been discovered, a notable example being topological states in quantum Hall systems~\cite{Klitzing1980,Tsui1982,Laughlin1983}. In the simplest cases, the integer quantum Hall (IQH) states can be understood as free electrons filling Landau levels. On the contrary, fractional quantum Hall (FQH) states only appear in strongly correlated systems. Fractionalized elementary excitations, multiple ground states on high-genus manifolds, and long-range quantum entanglement are their hallmarks.  The fact that quantum Hall states do not fit into the symmetry paradigm prompts the questions: what are the possible quantum phase transitions that involve quantum Hall states and how to characterize them? Previous works have investigated transitions between different IQH states~\cite{Chalker1988,Huckestein1990,HuoY1992,LeeDH1993}, between different FQH states~\cite{Jain1990-1,Kivelson1992,ZhuW2016-3,LeeJY2018}, and between certain IQH or FQH states and nontopological states~\cite{WenXG1993-2,YeJW1998,Barkeshli2014-2,Mulligan2010,Barkeshli2011,LiuZ2016,Motruk2017,ZhuW2019,ZhuZ2020,ZengTS2021,Kumar2022}.

The discovery of topological insulators greatly expanded the realm of topological phases~\cite{Hasan2010,QiXL2011-1}. One crucial insight of this adventure is that time-reversal and charge conservation symmetries should be preserved for these states to be nontrivial~\cite{Kane2005-1,Kane2005-2,Bernevig2006-1,Bernevig2006-2}. Further progress along this line leads to the concept of symmetry-protected topological (SPT) states~\cite{Schnyder2008,Kitaev2009,GuZC2009,Kitaev2011,ChenX2012,ChenX2013,Senthil2014}. This generalization incorporates strongly correlated states of spins, bosons, and fermions that exhibit nontrivial symmetry-protected edge physics but do not possess fractionalized excitations in the bulk. Quantum phase transitions from SPT states to trivial states or symmetry-breaking states have been studied~\cite{SonW2011,MizushimaT2012,Grover2013,XuCK2013,LuYM2014,Kshetrimayum2015,Scaffidi2016,YouYZ2016,WuHQ2016,Tsui2017,Scaffidi2017,Parker2018,ZengTS2020,XuYC2020,JianCM2021,Dupont2021-1,Dupont2021-2}.

In this work, we study a new class of topological phase transitions between SPT and FQH states in Bose-Fermi mixtures in the quantum Hall regime. We show that a SPT state can be realized for Bose-Fermi mixtures under suitable conditions, and it goes through a continuous transition to two decoupled FQH states as the interspecies interaction decreases. Experimentally, while fermionic quantum Hall states are routinely realized in solid state systems, bosonic ones are more challenging to realize~\cite{Gemelke2010,Aidels2013,TaiEM2017,Clark2019,Fletcher2021,Mukherjee2022,Leonard2023,ZhangDW2018,Cooper2019}. For cold atoms, Bose-Fermi mixtures have been extensively explored~\cite{Truscott2001,Modugno2002,Ferrier2014,YaoXC2016}. In solid state systems, electrons and holes may combine to form bosonic excitons. Electrons and excitons may coexist and form correlated Bose-Fermi mixtures in transition metal dichalcogenides~\cite{GaoBN2023,QiRS2023}. In addition, a recent work has reported evidence for the bosonic Laughlin state of excitons~\cite{WangR2023}. FQH states of excitons have also been proposed for moir\'e systems~\cite{Stefanidis2020,KwanYH2022}. This progress provides strong motivations for our investigations of FQH states and phase transitions in Bose-Fermi mixtures.

\begin{figure}
\includegraphics[width=0.45\textwidth]{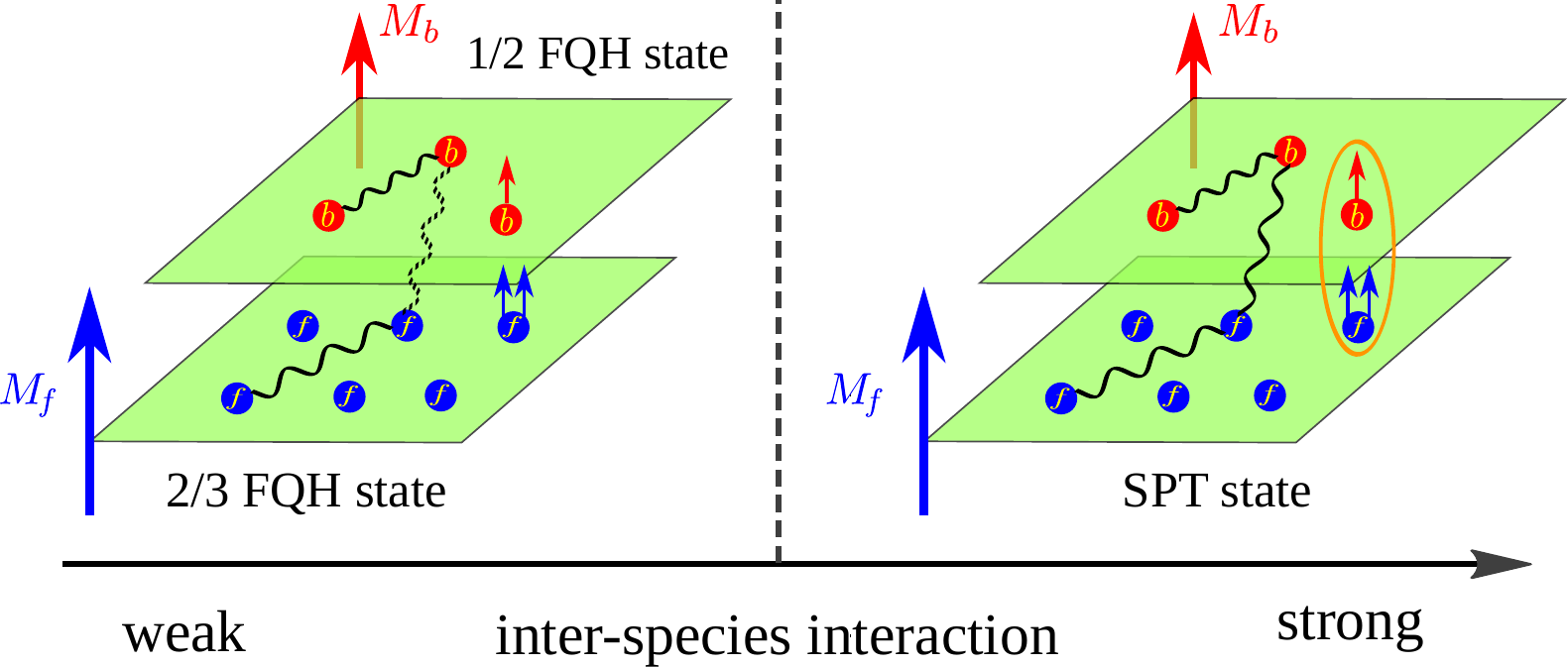}
\caption{Illustration of the quantum phase transition in Bose-Fermi mixtures. The solid (dashed) wiggle lines represent strong (weak) interactions between the particles. If there is no interspecies interaction, two independent FQH states are formed in which the particles are transformed to composite fermions (indicated by the small arrows). As the interspecies interaction strength grows, the two types of composite fermions eventually become strongly correlated to form the SPT state.}
\label{Figure1}
\end{figure}

{\em Wave functions and field theories for the SPT and FQH states} --- We start with trial wave functions for the SPT and FQH states in Landau levels, which will shine light on the nature of the transition. The letter $b$ ($f$) is used as subscripts or superscripts to represent bosons (fermions). For instance, the numbers of particles are denoted as $N_{b}$ and $N_{f}$. As illustrated in Fig.~\ref{Figure1}, the bosons and fermions are subjected to two independent magnetic fields with total fluxes $M_{b}$ and $M_{f}$, so their filling factors are $\nu_b=\frac{N_b}{M_b}$ and $\nu_f=\frac{N_f}{M_f}$. A positive direction for the magnetic fields is chosen so each filling factor has its sign. The IQH state with $\nu=n>0$ is denoted as $\Phi_{n}$ and that with $\nu=-n$ is $\Phi^{*}_{n}$. Throughout this work, we assume that the bosons/fermions carry a U(1) charge $e_{b}/e_{f}$. While in solid state systems one usually takes $e_{f}=1$ and $e_{b}$ an even integer (e.g. $e_{b}=2$ for Cooper pairs), this is not necessarily the case for cold atoms because they are actually charge neutral. Analogs of Hall conductance can be studied and the specific probing method determines the ``effective" charge of atoms~\cite{Leonard2023}.

In terms of the complex coordinates $z_{j},z_{k},\cdots$ on the plane, the SPT state is described by 
\begin{eqnarray}
\Psi_{\rm SPT} &\sim& \left[ \Phi^{*}_{1}(\{z^{b}_{j}\}) \Phi^{*}_{1}(\{z^{f}_{j}\}) \right] \prod^{N_{b}}_{j} \prod^{N_{f}}_{k} (z^{b}_{j}-z^{f}_{k}) \nonumber \\
        &\phantom{=}& \times \; \prod^{N_{b}}_{j<k} (z^{b}_{j}-z^{b}_{k}) \prod^{N_{f}}_{j<k} (z^{f}_{j}-z^{f}_{k})^{2} .
\label{eq:WaveFuncSPT}
\end{eqnarray}
It can be interpreted using the flux attachment process that maps strongly correlated particles to noninteracting composite fermions~\cite{Jain1989-1}: The bosons (fermions) are converted to composite fermions by the Jastrow factor $\prod^{N_{b}}_{j<k} (z^{b}_{j}-z^{b}_{k})$ $\left[ \prod^{N_{f}}_{j<k} (z^{f}_{j}-z^{f}_{k})^{2} \right]$; then the composite fermions form two $\nu=-1$ IQH states, and the interspecies correlation is captured by $\prod^{N_{b}}_{j} \prod^{N_{f}}_{k} (z^{b}_{j}-z^{f}_{k})$. In the thermodynamic limit, the numbers of particles and fluxes must satisfy $N_{b}=M_{f}$ and $N_{f}=M_{b}+M_{f}$ to realize $\Psi_{\rm SPT}$. In addition to the ground state, we can create four types of elementary excitations that carry integral charges~\cite{Supple}.

Topological properties of $\Psi_{\rm SPT}$ are encoded compactly in the Abelian Chern-Simons (CS) theory. The Lagrangian density is
\begin{eqnarray}
\mathcal{L}_{\rm CS} = \frac{1}{4\pi} K_{IJ} a_{I} d a_{J} + \frac{t_{I}}{2\pi} A d a_{I} \; ,
\end{eqnarray}
where $K$ is an integer-valued symmetric matrix, the $a_{I}$'s are emergent gauge fields, and $a_{I}da_{J} \equiv \epsilon^{\mu\nu\lambda} a_{I,\mu} \partial_{\nu} a_{J,\lambda}$. Here we also include the coupling with a background U(1) gauge field $A$, with integers $t_{I}$ known as the charge vector. This formalism was originally proposed for intrinsic topological orders~\cite{WenXG1992-1} but has also been very useful in studying SPT states~\cite{LuYM2012}. The number of degenerate ground states on a torus is given by $|\text{det}K|$. For the case with a unique ground state ($|\det K|=1$), one can further show that there exist no topologically nontrivial excitations.

Inspired by the wave function $\Psi_{\rm SPT}$, we consider the following $K$ matrix and charge vector:
\begin{eqnarray}
K_{\rm SPT} =
\begin{pmatrix}
0 & 1 \\
1 & 1
\end{pmatrix}, \quad 
\mb{t}_{\rm SPT} = 
\begin{pmatrix} 
e_{b} \\
e_{f}
\end{pmatrix}.
\end{eqnarray}
Because $K_{\rm SPT}$ has determinant $-1$ and zero signature (hence no chiral central charge), the theory indeed describes a SPT state. The Hall conductance of the system is $\sigma_{xy} = \mathbf{t}^{\mathsf{T}}_{\rm SPT} K_{\rm SPT}^{-1} \mathbf{t}_{\rm SPT} = e_{b}(2e_{f}-e_{b})$. If the system has an edge, there are two counterpropagating gapless modes with opposite chiralities, which can be protected by a U(1) symmetry when $\sigma_{xy}{\neq}0$ ($e_{b} \neq 0, 2e_{f}$). For the microscopic model studied below, the numbers of bosons and fermions are separately conserved, so we have a $\U_{b}{\times}\U_{f}$ symmetry. In this case, we can introduce two background gauge fields $A_{b}$ and $A_{f}$ that couple with the particles via
\begin{eqnarray}
\mb{t}^{b}_{\rm SPT} = 
\begin{pmatrix} 
e_{b} \\
0
\end{pmatrix}, \quad 
\mb{t}^{f}_{\rm SPT} = 
\begin{pmatrix} 
0 \\
e_{f}
\end{pmatrix}.
\end{eqnarray}
One can measure intraspecies Hall conductance (the response of one species to its associated field $A_{\sigma}$) and interspecies Hall conductance [the response of bosons to $A_{f}$ or fermions to $A_{b}$]. The results can be organized as a matrix
\begin{eqnarray}
\begin{pmatrix}
\sigma_{b} & \sigma_{\rm mix} \\
\sigma_{\rm mix} & \sigma_{f}
\end{pmatrix} = \begin{pmatrix}
-e^{2}_{b} & e_{b}e_{f} \\
e_{b}e_{f} & 0
\end{pmatrix}.
\label{ChernNumberMatrix}
\end{eqnarray}
In other words, the response theory contains a bosonic CS term $-\frac{1}{4\pi} e^{2}_{b} A_{b}dA_{b}$ and a mutual CS term $\frac{1}{2\pi} e_{b} e_{f} A_{b}dA_{f}$.

Now we turn to the FQH state in which the bosons and fermions are decoupled but still have suitable intraspecies interactions. At individual filling factors $\nu_{b}=1/2$ and $\nu_{f}=2/3$, the system is described by
\begin{eqnarray}
\Psi_{\rm FQH} &\sim& \Phi_{1}(\{z^{b}_{j}\}) \Phi^{*}_{2}(\{z^{f}_{j}\}) \nonumber \\
        &\phantom{=}& \times \; \prod^{N_{b}}_{j<k} (z^{b}_{j}-z^{b}_{k}) \prod^{N_{f}}_{j<k} (z^{f}_{j}-z^{f}_{k})^{2}.
\end{eqnarray}
Intuitively, the particles are also converted to composite fermions by the Jastrow factors, which now form their respective IQH states with $\nu=1$ (bosons) and $-2$ (fermions). In the CS theory, the bosonic FQH state has $K_{b}=2$ and $\mb{t}^{b}_{\rm FQH}=e_{b}$ and the fermionic FQH state has
\begin{eqnarray}
K_{f} = 
\begin{pmatrix}
1 & 0 \\
0 & -3
\end{pmatrix}, \quad 
\mb{t}^{f}_{\rm FQH} = e_{f} 
\begin{pmatrix}
1 \\
1
\end{pmatrix}.
\end{eqnarray}

{\em Quantum phase transitions} --- If we turn on interspecies interaction, it is possible to induce a quantum phase transition from the FQH state to the SPT state. To gain some intuition about how the transition takes place, we may strip off the flux attachment factors in $\Psi_{\rm SPT}$ and $\Psi_{\rm FQH}$ to consider a transition between the states $\Phi_{1}(\{z^{b}_{j}\}) \Phi^{*}_{2}(\{z^{f}_{j}\})$ and $\Phi^{*}_{1}(\{z^{b}_{j}\}) \Phi^{*}_{1}(\{z^{f}_{j}\}) \prod^{N_{b}}_{j} \prod^{N_{f}}_{k} (z^{b}_{j}-z^{f}_{k})$. The latter state is actually a superfluid because its $K$ matrix 
\begin{eqnarray}
\begin{pmatrix}
-1 & 1 \\
1 & -1
\end{pmatrix}
\end{eqnarray}
has zero determinant. This is reminiscent of the well-known exciton condensate in quantum Hall bilayers~\cite{Eisenstein2014}, but there the $K$ matrix has $1$ on the diagonal. In short, the transition may be understood as composite fermions change from two decoupled IQH states to one correlated superfluid.

This intuitive picture can be formalized using a field theory. It is helpful to perform a GL$(2,\mathbb{Z})$ transformation such that the $K$ matrix and charge vector for the fermionic state become
\begin{eqnarray}
K_{f} =
\begin{pmatrix}
1 & 1 \\
1 & -2
\end{pmatrix}, \quad
\mb{t}^{f}_{\rm FQH} = e_{f} 
\begin{pmatrix}
1 \\
0
\end{pmatrix}.
\end{eqnarray}
To combine the bosonic and fermionic FQH states, we rename the emergent gauge field for bosons as $a_{1}$ and the fields for fermions as $a_{2}$ and $a_{3}$. The resulting CS theory has $3{\times}3$-dimensional $K$ matrix $K_{\rm FQH} = K_{b}{\oplus}K_{f}$ and charge vector $\mb{t}_{\rm FQH} = \mb{t}^{b}_{\rm FQH} {\oplus} \mb{t}^{f}_{\rm FQH}$. Inspired by the analysis based on wave functions, we proceed to consider what happens when $a_{1}$ and $a_{3}$ are locked together by a Higgs field. Specifically, a complex scalar $\phi$ is introduced to construct the Lagrangian density
\begin{eqnarray}
\mathcal{L}_{\rm mix}  &=& \mathcal{L}_{b} + \mathcal{L}_{f} + \left| \left( \partial-ia_{1}+ia_{3} \right) \phi \right|^{2} \nonumber \\
             &\phantom{=}& + r|\phi|^{2} + u|\phi|^{4} + \cdots.
\end{eqnarray}
When $r>0$, $\phi$ is gapped and can be integrated out to reproduce the CS theory for the FQH state. When $r<0$, $\phi$ condenses to generate the Higgs phase in which $a_{3}$ can be eliminated by setting it to $a_{1}$. This leads to
\begin{eqnarray}
\frac{1}{2\pi}a_{1}da_{2} + \frac{1}{4\pi}a_{2}da_{2} + \frac{e_{b}}{2\pi}A_{b}da_{1} + \frac{e_{f}}{2\pi}A_{f}da_{2},
\end{eqnarray}
which is exactly the same as $\mathcal{L}_{\rm SPT}$. For a whole family of systems with filling factors $\nu_{b}=p/(p+1)$ and $\nu_{f}=(p+1)/(2p+1)$, we have uncovered similar mechanisms for continuous phase transitions and constructed the associated field theories~\cite{Supple}.

To further understand the critical theory, we perform the following GL(3,$\Z$) basis transformation for the gauge fields:
\begin{eqnarray}
\begin{pmatrix}
a_{1} \\
a_{2} \\
a_{3} 
\end{pmatrix}
= \begin{pmatrix}
3 & 1 & -1 \\
-2 & 0 & 1 \\
2 & 1 & -1
\end{pmatrix}
\begin{pmatrix}
b_{1} \\
b_{2} \\
b_{3}
\end{pmatrix}.
\end{eqnarray}
The $K$ matrix is
\begin{eqnarray}
\begin{pmatrix}
6 & 0 & 0 \\
0 & 0 & 1 \\
0 & 1 & -1
\end{pmatrix}
\end{eqnarray}
in the new basis. The critical theory becomes
\begin{eqnarray}
\frac{6}{4\pi} b_{1}db_{1} + |(\partial-ib_{1})\phi|^{2} + r|\phi|^{2} + u|\phi|^{4} + \cdots,
\label{eq:CriticalTheory}
\end{eqnarray}
so $a_{1}-a_{3}=b_{1}$ couples to $\phi$ while $b_{2}$ and $b_{3}$ decouple from critical fluctuations. Interestingly, this theory also describes a continuous transition between a $1/6$ Laughlin state and a trivial insulator. It is a strongly coupled theory for which analytical results are available only in the limit with a large number of boson flavors and a large CS level. In this case, (the generalization of) Eq.~\eqref{eq:CriticalTheory} indeed flows to a conformal fixed point at low energy. It is thus quite reasonable to conjecture that Eq.~\eqref{eq:CriticalTheory} describes an unconventional quantum critical point. For the transitions at other filling factors, similar basis transformations can also be found~\cite{Supple}.

\begin{figure}
\includegraphics[width=0.45\textwidth]{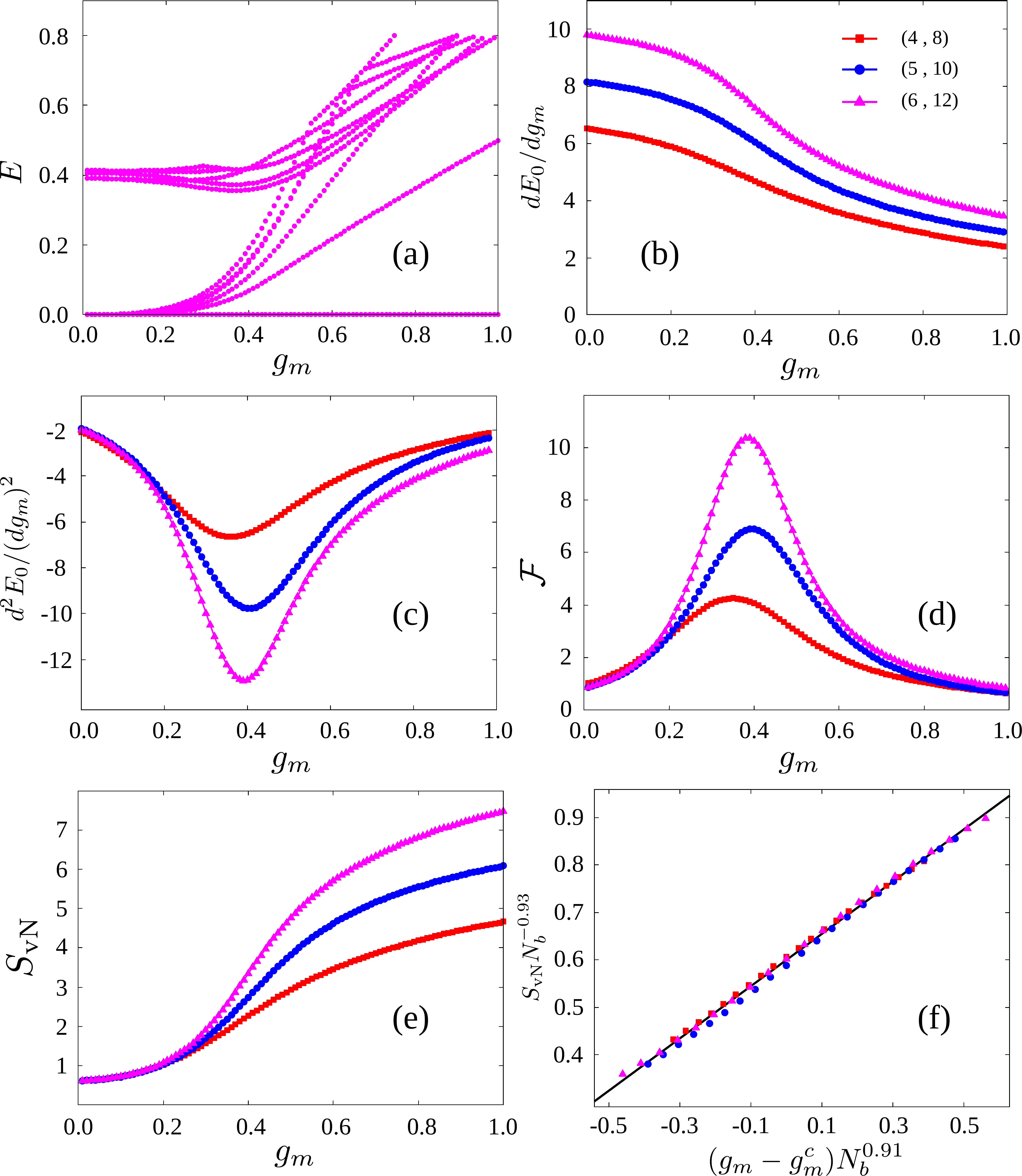}
\caption{Numerical results on the torus. (a) The low-lying energy levels of the $N_{b}=6,N_{f}=12$ system versus $g_{m}$. (b) The first-order derivative of the ground state energy. (c) The second-order derivative of the ground state energy. (d) The ground state fidelity susceptibility. (e) The von Neumann entanglement entropy between bosons and fermions. (f) The same data in (e) replotted to achieve data collapse. The numbers of particles $(N_{b},N_{f})$ for panels (b)-(f) are indicated using the legend of (b).}
\label{Figure2}
\end{figure}

{\em Numerical results} --- It is not {\it a priori} clear that the SPT state can be realized using a simple microscopic Hamiltonian. To this end, we consider the many-body Hamiltonian for the bosons and fermions
\begin{eqnarray}
H_{\rm mix} &=& \sum_{j<k} 4\pi\ell^{2}_{b} \; \delta( \mathbf{r}^{b}_{j} - \mathbf{r}^{b}_{k} ) + \sum_{j<k} 4\pi\ell^{4}_{f} \; \nabla^{2} \delta( \mathbf{r}^{f}_{j} - \mathbf{r}^{f}_{k} ) \nonumber \\
&+& g_{m} \sum_{j,k} 4\pi\ell_{b} \ell_{f} \; \delta( \mathbf{r}^{b}_{j} - \mathbf{r}^{f}_{k} ) \; ,
\end{eqnarray}
where $\ell_{b}$ ($\ell_{f}$) is the magnetic length for bosons (fermions). It is necessary to introduce two magnetic lengths because the magnetic fluxes for the two types of particles are different. The unit of length is chosen to be $\ell_{b}$. The particles are confined to their respective lowest Landau levels and higher levels are neglected. The first (second) term in $H_{\rm mix}$ corresponds to the zeroth (first) Haldane pseudopotential~\cite{Haldane1983-3}, so we know for sure that $\Psi_{\rm FQH}$ can be realized at $g_{m}=0$. Exact diagonalizations of $H_{\rm mix}$ are performed on the torus~\cite{Yoshioka1983} at many different $g_{m}\in[0,1]$. The energy spectra are presented in Fig.~\ref{Figure2} (a). A unique ground state is observed when $g_{m}{\sim}1$, but there are six quasidegenerate ground states when $g_{m}{\sim}0$~\cite{Haldane1985-2,WenXG1990}. This suggests that the Hamiltonian with $g_{m}{\sim}1$ resides in the SPT phase. To further corroborate this claim, we have computed the Hall conductance matrix for several cases using the twisted boundary condition method~\cite{NiuQ1985,FukuiT2005}. For simplicity, we choose $e_{b}=e_{f}=1$ in Eq.~\eqref{ChernNumberMatrix} such that the intraspecies Hall conductances are $-1$ and $0$ whereas the drag Hall conductance is $1$. The actual numbers obtained in numerical calculations are very close to these values (the deviation is smaller than $10^{-9}$)~\cite{Supple}.

The transition is inspected more closely using the lowest eigenvalue $E_{0}(g_{m})$ and the associated eigenstate $|\Psi_{0}(g_{m})\rangle$. The transition appears to be continuous, as one can see from the first-order derivative $dE_{0}/dg_{m}$ in Fig.~\ref{Figure2} (b). The transition point is found to be $g^{c}_{m} \approx 0.39$, where peaks appear in the second-order derivative $d^{2}E_{0}/dg_{m}^{2}$ as shown in Fig.~\ref{Figure2} (c). The evolution of $|\Psi_{0}(g_{m})\rangle$ can be characterized using the ground state fidelity susceptibility~\cite{Cozzini2007,YouWL2007}
\begin{eqnarray}
\mathcal{F}(g_{m}) = \frac{2}{ \left( \delta{g}_{m} \right)^{2} } \Big[ 1 - \big| \langle \Psi_{0}(g_{m}) | \Psi_{0}(g_{m}+\delta{g}_{m}) \rangle \big| \Big].
\end{eqnarray}
As the system passes the transition point, the state changes abruptly such that $\mathcal{F}$ attains a very large value. This picture is confirmed by the appearance of peaks around $g^{c}_{m} \approx 0.39$ in Fig.~\ref{Figure2} (d). The continuous nature of this transition is further corroborated by density matrix renormalization group calculations~\cite{Supple,White1992,Schollwock2011,HuZX2012}. In the vicinity of a critical point, critical scaling of physical quantities plays a prominent role. For symmetry-breaking phase transitions, correlation functions of local observables are routinely studied. However, they are not expected to give clear signatures due to the limited spatial extent of our system. To this end, we consider the quantum entanglement between the bosons and fermions. The reduced density matrix for the bosons is obtained by tracing out the fermions as $\rho_{b} = {\rm Tr}_{f} \; |\Psi_{0}(g_{m})\rangle\langle\Psi_{0}(g_{m})|$. The von Neumann entanglement entropy $S_{\rm vN} = -{\rm Tr} \; \rho_{b}\ln\rho_{b}$ is presented in Fig.~\ref{Figure2} (e). The boson-fermion entanglement is weak for small $g_{m}$ but seems to obey the volume law in the SPT state. Unfortunately, we are not able to derive the scaling form of $S_{\rm vN}$ using field theory. We make a bold conjecture that $S_{\rm vN}(g_{m})N^{\alpha}_{b}=f[(g_{m}-g^{c}_{m})N^{\beta}_{b}]$. The data points for $g_{m}\in[0.30,0.50]$ can be collapsed on a straight line using $\alpha\approx -0.93$ and $\beta\approx 0.91$ as shown in Fig.~\ref{Figure2} (f).

\begin{figure}
\includegraphics[width=0.45\textwidth]{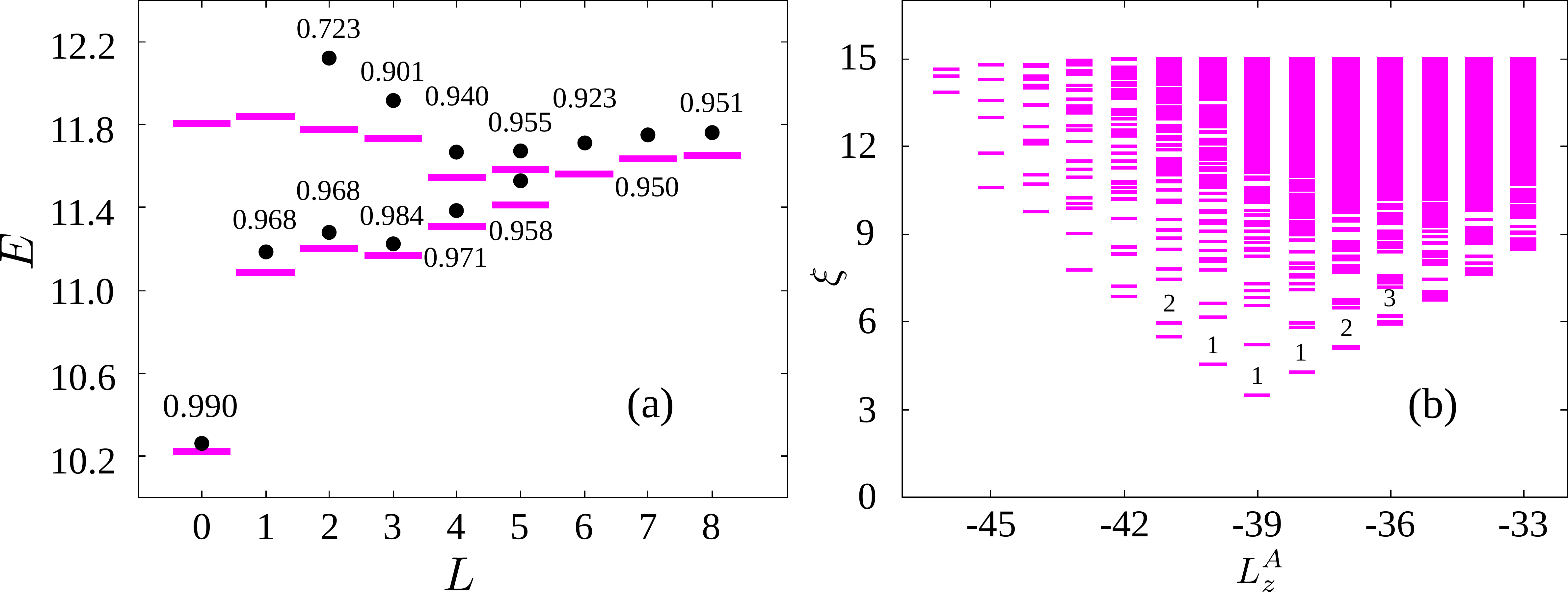}
\caption{Numerical results on the sphere. (a) The low-lying energy levels of the $N_{b}=5,N_{f}=8$ system. The lines (dots) represent exact eigenstates (trial wave functions) and the numbers are their overlaps. (b) The entanglement spectrum of the $N_{b}=7,N_{f}=12$ system in the sector for which the southern hemisphere has four bosons and six fermions.}
\label{Figure3}
\end{figure}

It is also helpful to employ the spherical geometry~\cite{Haldane1983-3}. A great advantage is that $\Psi_{\rm SPT}$ (and the trial wave functions for excitations) can be constructed more easily~\cite{Hermanns2013,PuSY2017}. However, its curvature results in a shift quantum number and the filling factor in finite-size systems may not be equal to its thermodynamic value~\cite{WenXG1992-2}. The system parameters should satisfy $M_{b}=2(N_{b}-1)$ for the bosonic $1/2$ state, $M_{f}=3N_{f}/2$ for the fermionic $2/3$ state, and $M_{b}=N_{f}, M_{f}=N_{b}+N_{f}$ for the SPT state. This imposes the condition $N_{b}=N_{f}/2+1$ instead of $N_{b}=N_{f}/2$. For the $N_{b}=5,N_{f}=8$ system, Fig.~\ref{Figure3} (a) displays the low-energy states of $H_{\rm mix}$ at $g_{m}=1.0$ (plotted versus the total angular momentum $L$), which are compared with appropriate trial wave functions~\cite{Supple}. The overlap for the ground state is excellent ($0.99$), and those for the excitations are quite good (except for one state). To probe the edge physics, we turn to the real space entanglement spectrum~\cite{LiH2008,Dubail2012-1,Sterdyniak2012,Rodriguez2012-2}. For the $N_{b}=7,N_{f}=14$ system, the eigenvalues of the reduced density matrix for the southern hemisphere are shown in Fig.~\ref{Figure3} (b). The good quantum numbers are the numbers of particles in the subspace and the $z$ component of the angular momentum. As indicated in the figure, two edge modes with opposite chiralities can be identified. The counting $1,1,2,3$ suggests that they are described by free bosons, which agrees with the CS theory.

{\em Conclusions} --- In summary, we have proposed an SPT state in Bose-Fermi mixtures that could be realized using a simple Hamiltonian. By tuning the interspecies interaction, quantum phase transitions to FQH states with intrinsic topological order can be induced. The possibility that these transitions are continuous is revealed by critical field theory and substantiated by numerical results. We have also made a first attempt toward revealing critical scaling of the entanglement entopy. This is very premature due to the absence of reliable analytical results on the scaling function. Many questions remain to be answered. It will be interesting to further study critical properties of the theory in Eq.~\eqref{eq:CriticalTheory}. More broadly, a general picture for the transitions between strongly correlated states in the quantum Hall regime is very desirable. The effects of disorder and other impefections that could appear in realistic systems should also be investigated. On the experimental frontier, multiple groups have reported FQH states in moir\'e systems without external magnetic field~\cite{CaiJQ2023,ZengYH2023,ParkHJ2023,XuF2023,LuZG2023}. A primitive idea is stacking these FQH states with the Laughlin state of excitons~\cite{WangR2023} to study phase transitions. Since Bose-Fermi mixtures have been realized in Refs.~\cite{GaoBN2023,QiRS2023} using electrons and excitons, it is natural to explore topological states in such systems.

{\em Note added} --- While finalizing the manuscript, we noticed a preprint on the transition between a FQH state and an exciton condensate in quantum Hall bilayers~\cite{ZhangYH2023}. The physics is quite different from the FQH-SPT transition studied in this work.

{\em Acknowledgements} --- We thank Chao-Ming Jian, Zhao Liu, Xin Wan, and Hao Wang for helpful conversations. This work was supported by the NNSF of China under grant No. 12174130 (Y.-H. W.), the Deutsche Forschungsgemeinschaft through project A06 of SFB 1143 under project No. 247310070 (H.-H. T.), and NSF under award No. DMR-1846109 (M.C.).

\bibliography{ReferConde}

\clearpage

\onecolumngrid

\setcounter{figure}{0}
\setcounter{equation}{0}
\renewcommand\thefigure{A\arabic{figure}}
\renewcommand\theequation{A\arabic{equation}}

\begin{appendix}

\section{Appendix A: Elementary excitations of the SPT state}

This section analyzes the elementary excitations of the symmetry-protected topological (SPT) state. Each boson (fermion) is attached with one (two) flux due to the same types of particles to become composite fermion. The interspecies correlation is accounted for by the Jastrow factor $\prod_{j<k} (z^{b}_{j}-z^{f}_{k})$. Alternatively, this factor may be interpreted as performing interspecies flux attachment such that each boson (fermion) is attached with one flux due to the other types of particles. The composite fermions reside in effective fluxes $M^{*}_{b}$ and $M^{*}_{f}$ and form their respective effective Landau levels (LLs). If the system parameters are chosen properly, the lowest LLs for the two types of composite fermions are completely filled to produce the $\nu=-1$ IQH states. This corresponds to the ground state of the physical system. If there are holes in the lowest effective LLs and/or particles in higher effective LLs, we would have excited states of the physical system.

\begin{figure}[ht]
\includegraphics[width=0.60\textwidth]{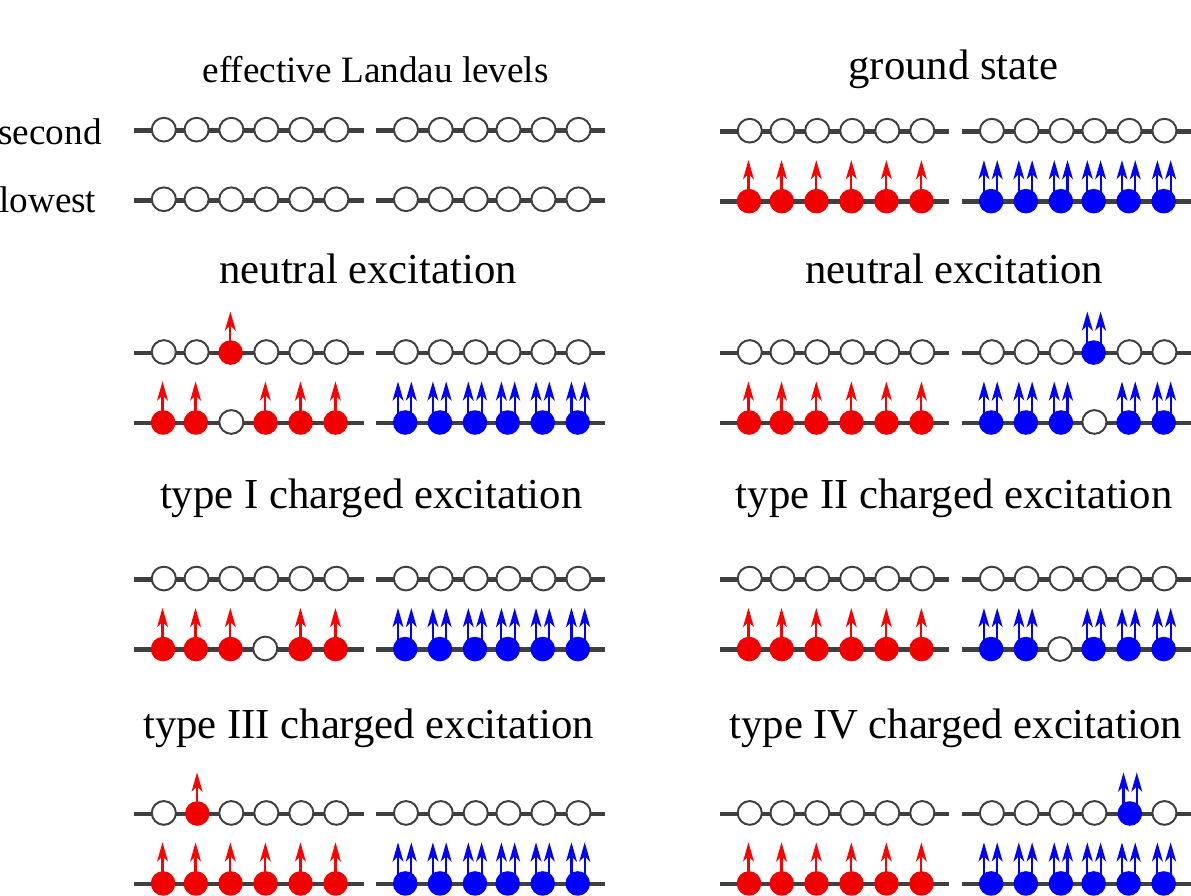}
\caption{Schematics of the elementary excitations of the SPT state.}
\label{FigureS1}
\end{figure}

We consider the elementary neutral and charged excitations shown in Fig.~\ref{FigureS1}. The latter name is borrowed from fractional quantum Hall (FQH) states, but it turns out that some of them may not carry charge. For the neutral excitations, the numbers of particles and the numbers of fluxes are the same as for the ground state, but one composite fermion is promoted from the fully occupied lowest effective LLs to the originally empty second effective LLs. For the charged excitations, there is only one hole in the lowest effective LLs or one particle in the second effective LLs. More specifically, the parameters should be adjusted as follows:
\begin{itemize}
\item type I: $N_{b},M_{b},M_{f}$ increase by one unit and $N_{f}$ is unchanged. 
\item type II: $N_{b}$ increases by one unit and $N_{f},M_{b},M_{f}$ are unchanged. 
\item type III: $N_{b},M_{b},M_{f}$ decrease by one unit and $N_{f}$ is unchanged. 
\item type IV: $N_{b}$ decreases by one unit and $N_{f},M_{b},M_{f}$ are unchanged. 
\end{itemize}
It is a little difficult to analyze their charges because the numbers of particles and fluxes are changed simultaneously. To solve this problem, we analyze the consequences of adding or removing one physical particle. If one boson is added, only one type II excitation is created, so we conclude that it has charge $e_{b}$. If one fermion is added, one type I and one type II excitations are created, so we conclude that type I excitation has charge $e_{f}-e_{b}$. Type III and type IV excitations are created if one particle is removed, and similar analysis shows that their charges are $e_{b}-e_{f}$ and $-e_{b}$, respectively.

\section{Appendix B: Phase transitions between Jain's series and SPT states}

This section generalizes the mechanism uncovered in the main text to a whole family of phase transitions between FQH and SPT states. On the FQH side, we have decoupled bosonic Jain state at $\nu_{b}=p/(p+1)$ and fermionic Jain state at $\nu_{f}=(p+1)/(2p+1)$ (with $p\in\mathbb{N}^{+})$. The SPT state is the same as in the main text, but the number of particles satisfy the constraint $N_{b}=pN_{f}/(p+1)$. The example studied in the main text corresponds to $p=1$.

We briefly review the $K$ matrix description of the FQH states in Jain's series. The bosonic Jain state at $\nu_{b}=p/(p+1)$ has $K$ matrix
\begin{eqnarray}
K_{b}=
\begin{pmatrix}
1 & 1 & \cdots & 1 \\
1 & 1 & \cdots & 1 \\
\vdots & \vdots & \ddots & \vdots \\
1 & 1 & \cdots & 1
\end{pmatrix} + \mathbb{I}_{p{\times}p}.
\end{eqnarray}
and charge vector is $\mb{t}^{b}_{\rm FQH}=e_{b}(1,1,\ldots,1)^\mathsf{T}$. Here $\mathbb{I}_{p{\times}p}$ is the $p{\times}p$ dimensional identity matrix. In this state, composite fermions are obtained by attaching one flux to each boson, and they fill $p$ Landau levels with Chern number $1$. The fermionic FQH state at $\nu_{f}=(p+1)/(2p+1)$ has $K$ matrix
\begin{eqnarray}
K_{f} = 2
\begin{pmatrix}
1 & 1 & \cdots & 1 \\
1 & 1 & \cdots & 1 \\
\vdots & \vdots & \ddots & \vdots \\
1 & 1 & \cdots & 1
\end{pmatrix} - \mathbb{I}_{(p+1){\times}(p+1)}.
\end{eqnarray}
and charge vector is $\mb{t}^{f}_{\rm FQH}=e_{f}(1,1,\ldots,1)^\mathsf{T}$. In this state, composite fermions are obtained by attaching two fluxes to each fermion, and they fill $p+1$ Landau levels with Chern number $-1$. 

When the two states are combined, the Chern-Simons theory would have $K_{\rm FQH}=K_{b}{\oplus}K_{f}$ and $\mb{t}_{\rm FQH} = \mb{t}^{b}_{\rm FQH}{\oplus}\mb{t}^{f}_{\rm FQH}$. An intuitive picture for the transition is that interspecies interaction induces strong correlation between composite fermions from the bosonic and fermionic Jain states. Formally, let us denote the gauge fields in the bosonic state as $a_{i}$ ($i=1,2,\ldots,p$) and those in the fermionic state as $a_{p+j}$ ($j=1,2,\ldots,p+1$). We propose that the transition is captured by the theory
\begin{eqnarray}
\mathcal{L}_{\rm mix} = \mathcal{L}_{b} + \mathcal{L}_{f} + \left| \left( \partial-ia_{p}+ia_{2p+1} \right) \phi \right|^{2}+ r|\phi|^{2} + u|\phi|^{4} + \cdots.
\end{eqnarray}
While no general proof has been found, it is very likely that the Higgsed phase is an invertible one and describes a SPT state. This claim can be verified easily for the $p=2$ case. In the Higgsed phase, the $K$ matrix takes the form
\begin{eqnarray}
K_{\rm SPT} =
\begin{pmatrix}
2 & 0 & 0 & 1 \\
0 & 1 & 2 & 2 \\
0 & 2 & 1 & 2 \\
1 & 2 & 2 & 3
\end{pmatrix}
\end{eqnarray}
and the charge vectors are $\mb{t}^{b}_{\rm SPT}=e_{b}(1,0,0,1)^\mathsf{T}, \quad \mb{t}^{f}_{\rm SPT}=e_{f}(0,1,1,1)^\mathsf{T}$. $K_{\rm SPT}$ has unity determinant and zero signature. This state and the one studied in the main text have the same Hall responses. While these features suggest that they are in the same SPT phase, their $K$ matrices do have different dimensions. To resolve this discrepancy, we change $K_\text{SPT}$ using a GL$(4,\mathbb{Z})$ transformation:
\begin{eqnarray}
\widetilde{K}_{\rm SPT} = W_{\rm SPT}^\mathsf{T} K_{\rm SPT} W_{\rm SPT} = 
\begin{pmatrix}
0 & 1 & 0 & 0 \\
1 & 0 & 0 & 0 \\
0 & 0 & 0 & 1 \\
0 & 0 & 1 & 1
\end{pmatrix}, \quad W_{\rm SPT} = 
\begin{pmatrix}
1 & 1 & -1 & 1 \\
0 & 1 & -1 & 1 \\
1 & 0 & -1 & 1 \\
-1 & -1 & 2 & -1
\end{pmatrix}.
\end{eqnarray}
The charge vectors in the new basis are 
\begin{eqnarray}
\widetilde{\mb{t}}^{b}_{\rm SPT} = e_{b}(0,0,1,0), \quad \widetilde{\mb{t}}^{f}_{\rm SPT} = e_{f}(0,0,0,1).
\end{eqnarray}
The upper-left $2{\times}2$ block $\begin{pmatrix} 0 & 1 \\ 1 & 0\end{pmatrix}$ is charge neutral and can be gapped out without breaking any symmetry. The lower-right $2{\times}2$ block and the associated components of the charge vectors are identical to those in the main text.

\begin{figure}[ht]
\includegraphics[width=0.80\textwidth]{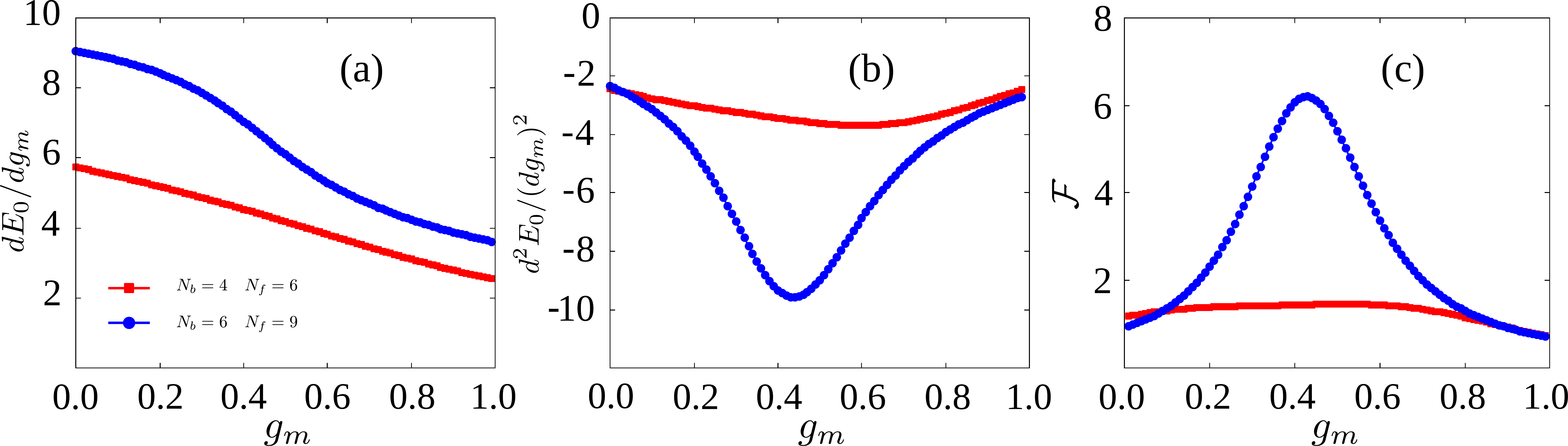}
\caption{Numerical results on the torus for the $p=2$ case. (a) The first-order derivative of the ground state energy. (b) The second-order derivative of the ground state energy. (c) The ground state fidelity susceptibility.}
\label{FigureS2}
\end{figure}

To better understand the critical theory, we can perform a ${\rm GL}(5,\mathbb{Z})$ transformation to rewrite $K_{\rm FQH}$ as
\begin{eqnarray}
\widetilde{K}_{\rm FQH}= W^{\mathsf{T}}_{2} K_{\rm FQH} W_{2} =
\begin{pmatrix}
15 & 0 & 0 & 0 & 0 \\
0 & -9 & 5 & 1 & 13 \\
0 & 5 & -1 & -1 & -4 \\
0 & 1 & -1 & 0 & -2 \\
0 & 13 & -4 & -2 & -14
\end{pmatrix}, \quad 
W_{2} = 
\begin{pmatrix}
-5 & -5 & 1 & 1 & 5 \\
10 & 11 & -2 & -2 & -10 \\
-6 & -6 & 2 & 1 & 6 \\
-6 & -6 & 1 & 1 & 6 \\
9 & 11 & -2 & -2 & -10
\end{pmatrix}.
\end{eqnarray}
The upper-left corner of $\widetilde{K}_{\rm FQH}$ decouples from the lower-right $4{\times}4$ block that has unity determinant and zero signature. In the transformed theory, we have gauge fields $b_{j}$ that are related to the $a_{j}$'s via $\mathbf{b}=W^{-1}_{2}\mathbf{a}$. The critical theory is simplified to
\begin{eqnarray}
\mathcal{L} = \frac{15}{4\pi} b_{1}db_{1} + |(\partial-ib_{1})\phi|^{2} + r|\phi|^{2} + u|\phi|^{4} + \cdots.
\end{eqnarray}
with $b_1=a_2-a_5$. Numerical calculations have been performed for the $p=2$ case and the results are presented in Fig.~\ref{FigureS2}. It is plausible that the transition is continuous, but finite-size effect is quite strong here.

\section{Appendix C: Details about the Hall conductance matrix}

This section explains in detail how to compute the Hall conductance matrix. The system is placed on a rectangular torus whose two sides have lengths $L_{x}$ and $L_{y}$. The unit vectors along the two directions of the torus are denoted as $\mathbf{e}_{x}$ and $\mathbf{e}_{y}$. The reciprocal vectors are 
\begin{eqnarray}
\mathbf{G}_{1} = \frac{2\pi}{L_{x}} \mathbf{e}_{x}, \qquad \mathbf{G}_{2} = \frac{2\pi}{L_{y}} \mathbf{e}_{y}.
\end{eqnarray}
The translations along $L_{x}\mathbf{e}_{x}$ and $L_{y}\mathbf{e}_{y}$ must commute. The magnetic fluxes $N^{\sigma}_{\phi}$ for the two layers are constrained by the equation
\begin{eqnarray}
2\pi\ell^{2}_{\sigma}N^{\sigma}_{\phi} = L_{x}L_{y}.
\end{eqnarray}
The many-body Chern numbers are computed by applying twisted boundary conditions in the system~\cite{NiuQ1985}. For our bilayer system, there are four boundary twist angles $\theta^{\sigma}_{x}$ and $\theta^{\sigma}_{y}$. In the presence of these angles, the single-particle eigenstates are
\begin{eqnarray}
\phi^{\sigma}_{m}(x,y,\theta_{x},\theta_{y}) &=& \frac{1}{ ( \sqrt{\pi}\ell_{\sigma}L_{y} )^{1/2}} \sum^{\mathbb{Z}}_{k} \exp \left( ik\theta_{x} \right) \nonumber \\
&\times& \exp \left\{ -\frac{1}{2} \left[ \frac{x}{\ell_{B}} - \frac{2{\pi}\ell_{\sigma}}{L_{y}} \left( \frac{\theta_{y}}{2\pi} + m + kN^{\sigma}_{\phi} \right) \right]^{2} + i \frac{2{\pi}y}{L_{y}} \left( \frac{\theta_{y}}{2\pi} + m + kN^{\sigma}_{\phi} \right) \right\}.
\end{eqnarray}
The creation (annihilation) operator associated with $\phi^{\sigma}_{m}(x,y,\theta_{x},\theta_{y})$ is denoted as $C^{\dag}_{\sigma,m}$ ($C_{\sigma,m}$).

For a general two-body potential $V^{\sigma\tau}(\mathbf{r}_{1}-\mathbf{r}_{2})$, the second quantized form is
\begin{eqnarray}
H = \frac{1}{2} \sum_{\sigma,\tau} \sum_{\{m_{i}\}} V^{\sigma\tau}_{m_{1}m_{2}m_{3}m_{4}} C^{\dag}_{\sigma,m_{1}} C^{\dag}_{\tau,m_{2}} C_{\tau,m_{4}} C_{\sigma,m_{3}}
\end{eqnarray}
with
\begin{eqnarray}
V^{\sigma\tau}_{m_{1}m_{2}m_{3}m_{4}} = \int d \mathbf{r}_{1} \int d\mathbf{r}_{2} \left[ \phi^{\sigma}_{m_{1}}(\mathbf{r}_{1}) \right]^{*} \left[ \phi^{\tau}_{m_{2}}(\mathbf{r}_{2}) \right]^{*} \; V^{\sigma\tau}(\mathbf{r}_{1}-\mathbf{r}_{2}) \; \phi^{\tau}_{m_{4}}(\mathbf{r}_{2}) \phi^{\sigma}_{m_{3}}(\mathbf{r}_{1}).
\end{eqnarray}
It is convenient to perform Fourier transform on $V^{\sigma\tau}(\mathbf{r}_{1}-\mathbf{r}_{2})$ such that the Hamiltonian becomes
\begin{eqnarray}
H = \frac{1}{2L_{x}L_{y}} \sum_{\mathbf{q}} V^{\sigma\tau}(\mathbf{q}) : \rho^{\sigma}(\mathbf{q}) \rho^{\tau}(-\mathbf{q}) :
\end{eqnarray}
and
\begin{eqnarray}
&& \rho^{\sigma}(\mathbf{q}) = \int d \mathbf{r}_{1} \sum_{m_{1},m_{3}} \left[ \phi^{\sigma}_{m_{1}}(\mathbf{r}_{1}) \right]^{*} \exp \left( i\mathbf{q}\cdot\mathbf{r}_{1} \right) \phi^{\sigma}_{m_{3}}(\mathbf{r}_{1}) \; C^{\dag}_{\sigma,m_{1}} C_{\sigma,m_{3}}, \nonumber \\
&& \rho^{\tau}(-\mathbf{q}) = \int d \mathbf{r}_{2} \sum_{m_{2},m_{4}} \left[ \phi^{\tau}_{m_{2}}(\mathbf{r}_{2}) \right]^{*} \exp \left( -i\mathbf{q}\cdot\mathbf{r}_{2} \right) \phi^{\tau}_{m_{4}}(\mathbf{r}_{2}) \; C^{\dag}_{\tau,m_{2}} C_{\tau,m_{4}}.
\end{eqnarray}
The integrals can be evaluated to yield
\begin{eqnarray}
H &=& \frac{1}{2L_{x}L_{y}} \sum_{\sigma,\tau} \sum_{\{m_{i}\}} \sum_{q_{1},q_{2}} V(\mathbf{q}) \exp \left\{ -\frac{1}{4} |\mathbf{q}|^{2} \left( \ell^{2}_{\sigma} + \ell^{2}_{\tau} \right) \right\} \nonumber \\
&\times& \exp \left[ i\frac{\theta^{\sigma}_{x}}{N^{\sigma}_{\phi}} (m_{1}-m_{3}-q_{2}) \right] \exp \left[ i\frac{\theta^{\tau}_{x}}{N^{\tau}_{\phi}} (m_{2}-m_{4}+q_{2}) \right] \nonumber \\
&\times& \exp \left[ i\frac{{\pi}q_{1}}{N^{\sigma}_{\phi}} \left( \frac{\theta^{\sigma}_{y}}{\pi} + 2m_{1}-q_{2} \right) - i\frac{{\pi}q_{1}}{N^{\tau}_{\phi}} \left( \frac{\theta^{\tau}_{y}}{\pi} + 2m_{2}+q_{2} \right) \right] \nonumber \\
&\times& {\widetilde\delta}_{m_{1},m_{3}+q_{2}} {\widetilde\delta}_{m_{2},m_{4}-q_{2}} C^{\dag}_{\sigma,m_{1}} C^{\dag}_{\tau,m_{2}} C_{\tau,m_{4}} C_{\sigma,m_{3}} 
\end{eqnarray}
where we have introduced the generalized Kronecker symbol
\begin{eqnarray}
\widetilde{\delta}_{s,t+q_{2}} = 1 \qquad \text{if and only if} \qquad s \; \text{mod} \; N_{\phi} = (t+q_{2}) \; \text{mod} \; N_{\phi}.
\end{eqnarray}

For a given set of boundary twist angles, the SPT phase has a unique ground state that we denote as $|\Phi(\theta^{b}_{x},\theta^{b}_{y},\theta^{f}_{x},\theta^{f}_{y})\rangle$. We perform another unitary transformation to define the state
\begin{eqnarray}
|\Psi(\theta^{b}_{x},\theta^{b}_{y},\theta^{f}_{x},\theta^{f}_{y}) \rangle = \exp \left[ -i \sum_{\sigma=b,f} \sum^{N}_{k=1} \left( \frac{x_{k}}{L_{x}} \theta^{\sigma}_{x} + \frac{y_{k}}{L_{y}} \theta^{\sigma}_{y} \right) \right] |\Phi(\theta^{b}_{x},\theta^{b}_{y},\theta^{f}_{x},\theta^{f}_{y}) \rangle.
\label{ChernUnitaryTrans}
\end{eqnarray}
The exponential can be decomposed to be a products of exponentials for each particle. The elements of the Hall conductance matrix are the Chern numbers
\begin{eqnarray}
C_{\sigma\tau} = \frac{1}{2{\pi}i} \int^{2\pi}_{0} d\theta^{\sigma}_{x} \int^{2\pi}_{0} d\theta^{\tau}_{y} \left[ \left\langle \frac{\partial\Psi}{\partial\theta^{\sigma}_{x}} \left| \frac{\partial\Psi}{\partial\theta^{\tau}_{y}} \right. \right\rangle - \left\langle \frac{\partial\Psi}{\partial\theta^{\tau}_{y}} \left| \frac{\partial\Psi}{\partial\theta^{\sigma}_{x}} \right. \right\rangle \right].
\end{eqnarray}
In our calculation, the Wilson loop method is employed since it is more accurate than direct numerical integration~\cite{FukuiT2005}. Let us explain the procedure in detail using $C_{bb}$. No boundary twist angles are applied to the fermions so they are dropped from the formulas to avoid clutter. The interval $[0,2\pi]$ is divided into $S$ segments such that each segment has length $\Delta\theta=2\pi/S$. The many-body Hamiltonian with $\theta^{b}_{x}=i\Delta\theta$ and $\theta^{b}_{y}=j\Delta\theta$ is diagonalized to generate the state $|\Psi(i,j) \rangle$. The non-unitary exponential Berry connections are defined as
\begin{eqnarray}
\mathcal{A}^{bb}_{x}(i,j) = \langle \Psi(i,j) | \Psi(i+1,j) \rangle, \qquad \mathcal{A}^{bb}_{y}(i,j) = \langle \Psi(i,j) | \Psi(i,j+1) \rangle.
\label{NonUnitaryBerry}
\end{eqnarray}
This leads to the unitary Berry connections
\begin{eqnarray}
A^{bb}_{x}(i,j) = \frac{\mathcal{A}^{bb}_{x}(i,j)}{|\mathcal{A}^{bb}_{x}(i,j)|}, \qquad A^{bb}_{y}(i,j) = \frac{\mathcal{A}^{bb}_{y}(i,j)}{|\mathcal{A}^{bb}_{y}(i,j)|}.
\end{eqnarray}
For the plaquette whose lower-left corner is located at $i,j$, its U(1) Wilson loop is defined as
\begin{eqnarray}
W_{bb}(i,j) = A^{bb}_{x}(i,j) A^{bb}_{y}(i+1,j) \left[ A^{bb}_{x}(i,j+1) \right]^{*} \left[ A^{bb}_{y}(i,j) \right]^{*}.
\end{eqnarray}
The Chern number is
\begin{eqnarray}
C_{bb} = \frac{1}{2\pi} \sum^{S-1}_{i,j=0} \Im \log \left[ W_{bb}(i,j) \right],
\end{eqnarray}
where the imaginary part is restricted to the range $(-\pi,\pi]$. For the Chern number $C_{\sigma\tau}$, we only need to use nonzero $\theta^{\sigma}_{x}$ and $\theta^{\tau}_{y}$ and the procedure is similar.

The essential step in this method is computing the overlaps in Eq.~\eqref{NonUnitaryBerry}. It is important to recognize that the Fock states depend on the boundary twist angles, and we need to know the overlaps between the single-particle eigenstates with different angles. The unitary transformation in Eq.~\eqref{ChernUnitaryTrans} can be implemented by changing the single-particle eigenstates to
\begin{eqnarray}
\widetilde{\phi}^{\sigma}_{m}(x,y,\theta^{\sigma}_{x},\theta^{\sigma}_{y}) &=& \exp\left( -i\frac{\theta^{\sigma}_{x}x}{L_{x}} - i\frac{\theta^{\sigma}_{y}y}{L_{y}} \right) \frac{1}{ ( \sqrt{\pi}\ell_{\sigma}L_{y} )^{1/2}} \sum^{\mathbb{Z}}_{k} \exp \left( ik\theta^{\sigma}_{x} \right) \nonumber \\
&\phantom{=}& \times \exp \left\{ -\frac{1}{2} \left[ \frac{x}{\ell_{\sigma}} - \frac{2{\pi}\ell_{\sigma}}{L_{y}} \left( \frac{\theta_{y}}{2\pi} + m + kN^{\sigma}_{\phi} \right) \right]^{2} + i \frac{2{\pi}y}{L_{y}} \left( \frac{\theta_{y}}{2\pi} + m + kN^{\sigma}_{\phi} \right) \right\} \nonumber \\
                                    &=& \exp\left( -i\frac{\theta^{\sigma}_{x}x}{L_{x}} \right) \frac{1}{ ( \sqrt{\pi}\ell_{B}L_{y} )^{1/2}} \sum^{\mathbb{Z}}_{k} \exp \left( ik\theta^{\sigma}_{x} \right) \nonumber \\
&\phantom{=}& \times \exp \left\{ -\frac{1}{2} \left[ \frac{x}{\ell_{\sigma}} - \frac{2{\pi}\ell_{\sigma}}{L_{y}} \left( \frac{\theta^{\sigma}_{y}}{2\pi} + m + kN^{\sigma}_{\phi} \right) \right]^{2} + i \frac{2{\pi}y}{L_{y}} \left( m + kN^{\sigma}_{\phi} \right) \right\}
\end{eqnarray}
For two states with the same $\theta^{\sigma}_{x}$, we have
\begin{eqnarray}
\int d\mathbf{r} \left[ \widetilde{\phi}^{\sigma}_{m_{1}}(x,y,\theta^{\sigma}_{x},\theta^{\sigma}_{y}) \right]^{*} \widetilde{\phi}^{\sigma}_{m_{2}}(x,y,\theta^{\sigma}_{x},\zeta^{\sigma}_{y}) = \delta_{m_{1},m_{2}} \exp \left[ - \frac{\ell^{2}_{\sigma}}{4L^{2}_{y}} (\zeta^{\sigma}_{y}-\theta^{\sigma}_{y})^{2} \right].
\end{eqnarray}
For two states with the same $\theta^{\sigma}_{y}$, we have
\begin{eqnarray}
\int d \mathbf{r} \left[ \widetilde{\phi}^{\sigma}_{m_{1}}(x,y,\theta^{\sigma}_{x},\theta^{\sigma}_{y}) \right]^{*} \widetilde{\phi}^{\sigma}_{m_{2}}(x,y,\zeta^{\sigma}_{x},\theta^{\sigma}_{y}) = \delta_{m_{1},m_{2}} \exp \left[ - \frac{\ell^{2}_{\sigma}}{4L^{2}_{x}} (\zeta^{\sigma}_{x}-\theta^{\sigma}_{x})^{2} -\frac{i}{N^{\sigma}_{\phi}} \left( \frac{\theta^{\sigma}_{y}}{2\pi} + m_{1} \right) (\zeta^{\sigma}_{x}-\theta^{\sigma}_{x}) \right].
\end{eqnarray}
In our calculations, the interval $[0,2\pi]$ is divided into 20 segments. The Hall conductance matrix has been computed in many cases. For the SPT phase, only the absolute ground state is needed for this calculation. The matrix at $g_{m}=1.0$ is
\begin{eqnarray}
\begin{pmatrix}
-1 & 1 \\
1 & 0
\end{pmatrix} + 
\begin{pmatrix}
1.96{\times}10^{-10} & 3.79{\times}10^{-11} \\
2.44{\times}10^{-10} & 3.07{\times}10^{-10}
\end{pmatrix} 
\end{eqnarray}
for $N_{b}=4,N_{f}=8$ and 
\begin{eqnarray}
\begin{pmatrix}
-1 & 1 \\
1 & 0
\end{pmatrix} +
\begin{pmatrix}
-4.06{\times}10^{-10} & 3.08{\times}10^{-10} \\
6.43{\times}10^{-11} & -1.72{\times}10^{-10}
\end{pmatrix} 
\end{eqnarray}
for $N_{b}=5,N_{f}=10$. These agree with Eq.~(5) in the main text. For the FQH phase, all six quasi-degenerate ground states should be included. The drag Hall conductance at $g_{m}=0.2$ is $-6.01{\times}10^{-10}$ for $N_{b}=4,N_{f}=8$ and $-2.56{\times}10^{-9}$ for $N_{b}=5,N_{f}=10$. This confirms that the two species are independent.

\section{Appendix D: Additional numerical results}

\begin{figure}[ht]
\includegraphics[width=0.50\textwidth]{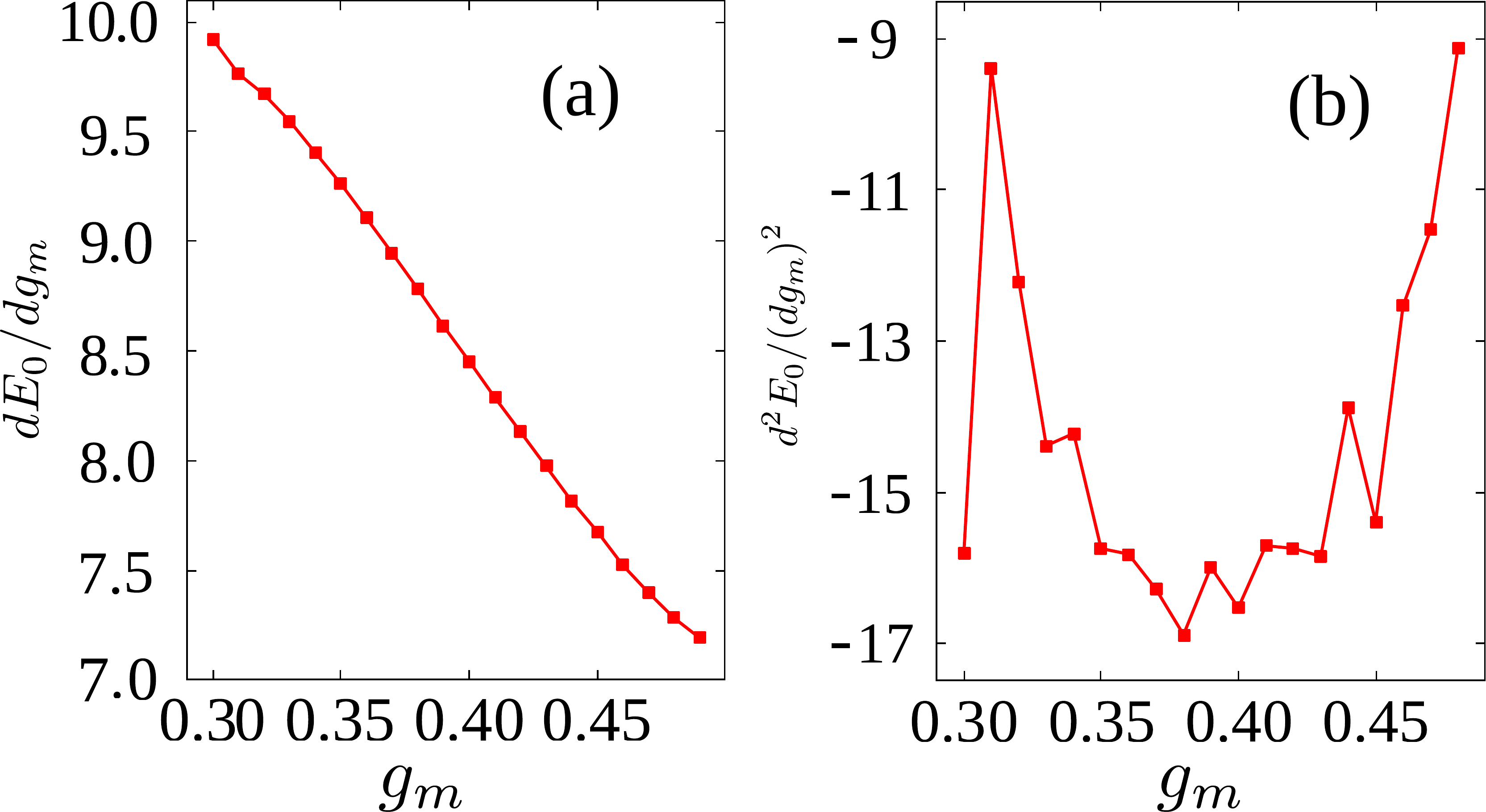}
\caption{DMRG results for the $N_{b}=7, N_{f}=14$ system on the torus. (a) The first-order derivative of the ground state energy. (b) The second-order derivative of the ground state energy.}
\label{FigureS3}
\end{figure}

The system size that can be accessed in exact diagonalization is limited by the exponential growth of the Hilbert space dimension. In many cases, it is possible to study larger systems using the density matrix renormalization group (DMRG), which searches for the ground state in the manifold of matrix product states~\cite{White1992,Schollwock2011}. DMRG calculations have been performed using the Hamiltonian $H_{\rm mix}$ on the torus, but the results are not particularly successful. For the $N_{b}=7, N_{f}=14$ system, the first-order and second-order derivatives of the ground state energy $E_{0}(g_{m})$ are presented in Fig.~\ref{FigureS3}. The former curve is quite smooth but the latter curve has strong fluctuations. The first-order derivative is approximated by the difference
\begin{eqnarray}
\frac{E_{0}(g_{m}+\delta{g}_{m})-E_{0}(g_{m})}{\delta{g}_{m}}.
\end{eqnarray}
To obtain accurate values with $\delta{g}_{m}=0.01$, the uncertainty of $E_{0}$ should be less than $O(10^{-3})$, which is somewhat challenging but can be done in a reasonable amount of time. For the second-order derivative, accurate values can be obtained only if the uncertainty of $E_{0}$ is less than $O(10^{-5})$, which is too demanding and not achieved in our calculations. This is not surprising in view of the fundamental difficulty on the torus~\cite{HuZX2012}. 

\end{appendix}

\end{document}